\documentclass[%
 reprint,
 amssymb, amsmath,%
 aip,cha,
]{revtex4-1}

\usepackage{bm}%
\usepackage[colorlinks=true,linkcolor=blue]{hyperref}%
\usepackage{graphicx}
\expandafter\ifx\csname package@font\endcsname\relax\else
 \expandafter\expandafter
 \expandafter\usepackage
 \expandafter\expandafter
 \expandafter{\csname package@font\endcsname}%
\fi
\begin{document}

\title{Mitigation of dynamical instabilities  in laser arrays via non-Hermitian coupling}%

\author{S. Longhi}%
\email{longhi@fisi.polimi.it}
\affiliation{Dipartimento di Fisica, Politecnico di Milano and Istituto di Fotonica e Nanotecnologie del Consiglio Nazionale delle Ricerche, Piazza L. da Vinci 32, I-20133 Milano, Italy}%

\author{L. Feng}%
%\email{}
\affiliation{Department of Electrical and Systems Engineering, University of Pennsylvania, Philadelphia, PA 19104, USA}%

\date{March 2018}%
%\revised{August 2010}%

\begin{abstract}
\noindent  
Arrays of coupled semiconductor lasers are systems possessing complex dynamical behavior that are of major interest in photonics and laser science. Dynamical instabilities, arising from supermode competition and slow carrier dynamics, are known to prevent stable phase locking in a wide range of parameter space, requiring special methods to realize stable laser operation. Inspired by recent concepts of parity-time ($\mathcal{PT}$) and non-Hermitian photonics, in this work we consider non-Hermitian coupling engineering in laser arrays in a ring geometry and show, both analytically and numerically,  that non-Hermitian coupling can help to mitigate the onset of dynamical laser instabilities. In particular, we consider in details two kinds of nearest-neighbor non-Hermitian couplings: symmetric but complex mode coupling (type-I non-Hermitian coupling) and asymmetric mode coupling (type-II non-Hermitian coupling). Suppression of dynamical instabilities can be realized in both coupling schemes, resulting in stable phase-locking laser emission with the lasers emitting in phase (for type-I coupling) or with $\pi/2$ phase gradient (for type-II coupling), resulting in a vortex far-field beam. In type-II non-Hermitian coupling, chirality induced by asymmetric mode coupling enables laser phase locking even in presence of moderate disorder in the resonance frequencies of the lasers. 
\end{abstract}

\maketitle

%\tableofcontents

\section{Introduction}
Non-Hermitian and parity-time ($\mathcal{PT}$) symmetric photonics, i.e. the ability of molding the flow of light in synthetic optical media by  judicious spatial distribution of optical gain and loss, is an emerging and active area of research in optics (see e.g. \cite{r1,r2,r3,r4,r4bis} and references therein). Inspired by concepts of non-Hermitian quantum mechanics \cite{r5,r6,r7,r8} and originally conceived to provide an experimentally accessible testbed to emulate in optics  non-Hermitian scattering potentials and quantum phase transitions \cite{r9,r10,r11,r12,r13,r14,r15,r15bis,r16}, $\mathcal{PT}$ symmetric photonics has demonstrated to be a fertile and technologically accessible research field which is promising for a wealth of interesting applications \cite{r18,r19,r20,r21,r22,r23,r24,r25,r26,r27,r28,r29,r30,r31,r32,r33,r34,r35,r36,r37,r38,r39,r40,r41,r42,r43,r44,r45,r46} ranging from material transparency and invisibility \cite{r20,r21,r22,r23,r24}, laser-absorber devices \cite{r18,r19,r28,r32}, microlaser engineering and mode selection \cite{r25,r26,r27,r29,r31,r35,r37}, polarization mode conversion \cite{r38}, light structuring and transport \cite{r30,r33}, optical sensing \cite{r39,r39bis,r40,r41}, topological lasers \cite{r44,r45,r46}, etc. The application of the concepts of non-Hermitian optics in integrated laser devices, i.e. beyond linear models, meets the problem of complexity and nonlinear instabilities typical of laser systems \cite{r47,r48,r49}. $\mathcal{PT}$ symmetry and non-Hermitian engineering have recently emerged as useful tools in the control of laser dynamics \cite{r25,r26,r27,r29,r30,r31,r35,r36,r37,r42,r43,r45,r46}, including systems of coupled laser arrays \cite{r36,r42,r43,r44,r45}, and in laser mode locking \cite{r49bis}.  \par
 Stable oscillation of arrays of coupled lasers in a given supermode is a longstanding problem in laser science and technology  \cite{r50,r51,r52,r53,r54,r55,r56,r57,r58,r59,r60,r61,r62,r63,r64,r65,r66,r66bis,r67,r67bis,r68,r68bis,r69,r70,r70bis}. Avoiding instabilities is of great technological
importance for the realization of high-power laser arrays and for a variety of applications in optical communications,
sensing, and imaging \cite{r50,r51,r52,r53}. Unfortunately, stable phase-locked oscillation in laser arrays is often prevented  by supermode competition and laser instabilities\cite{r55,r58,r61,r62,r64,r65,r66,r67bis,r69,r70bis}:  the complicated array dynamics can lead to unstable behavior in a wide range of physically meaningful parameter space. Careful laser design, based on gain tailoring and/or special diffractive coupling, is hence needed to achieve stable phase locking operation \cite{r54,r56,r57,r59,r60,r63,r68,r68bis}. In particular, in semiconductor lasers the slow carrier dynamics and the large linewidth enhancement factor severely narrow the parameter space region of  stable phase locking laser operation.\par
 In this article we apply concepts of non-Hermitian photonics to the control of the dynamical behavior of coupled semiconductor lasers in a  ring geometry, and show both analytically and numerically that nearest-neighbor non-Hermitian coupling engineering can help in suppressing the onset of dynamical instabilities. In particular, we consider in details two kinds of nearest-neighbor non-Hermitian couplings, so called type-I and type-II non-Hermitian couplings. We show that suppression of dynamical instability can be realized, resulting in stable phase-locking laser emission with the lasers emitting with the same phase (for type-I non-Hermitian coupling) or with $\pi/2$ phase slip one another (for type-II non-Hermitian coupling). In the latter case a vortex far-field beam can be achieved. The paper is organized as follows. Section II describes the rate equations model for coupled semiconductor lasers in a ring geometry and with rather general global or local (nearest-neighbor) non-Hermitian coupling, and presents simple phase-locked stationary states under a few coupling schemes. The stability analysis of the phase-locked solutions is presented in Sec. III, where analytical stability boundaries are derived using an asymptotic method. In particular, it is shown that appropriate tailoring of non-Hermitian neighboring couplings can lead to suppression of dynamical (Hopf) instability generally observed when dealing with Hermitian coupling \cite{r64}. In Sec. IV some numerical results are presented, which confirm the predictions of the theoretical analysis. Finally, the main conclusions are summarized in Sec.V.
 
\section{Semiconductor laser arrays with non-Hermitian coupling}
\subsection{Rate equations model}
We consider an array of $N$ semiconductor lasers in a ring geometry \cite{r61,r64,r70,r71,r72,r73},  schematically depicted in Fig.1(a), which are locally or globally coupled by either evanescent mode coupling or by some diffractive coupling technique (see, for instance, \cite{r51, r54,r57,r59,r60,r63,r67,r68,r68bis,r70,r71,r72,r73} and references therein). The rate equations that describe the temporal evolution of the slowly varying complex amplitudes of normalized
electric fields $E_n$ and normalized excess carrier density $Z_n$ in each laser read \cite{r58,r62,r64,r69}
\begin{eqnarray}
\frac{dE_n}{dt} & = & (1-i \alpha)Z_nE_n-i \sum_{l=1}^{N} \kappa_{n,l}E_l \\
T \frac{dZ_n}{d t} & = & p-Z_n-(1+2Z_n)|E_n|^2
\end{eqnarray}
($n=1,2,...,N$), where $t$ is dimensionless time in units of the photon lifetime $\tau_p$, $\alpha$ is the linewidth-enhancement factor (typically $ \alpha \simeq 3-5$) , $p$ is the normalized excess pump
current, $T= \tau_s / \tau_p$ is the ratio between the spontaneous carrier lifetime $\tau_s$ and the photon lifetime $\tau_p$ (typically in the range $T \sim 100-1000$), and the matrix $\kappa_{n,l}$ describes the coupling between the various lasers in the array. As in Refs.\cite{r62,r64}, in writing Eqs.(1) and (2) we neglected time-delay effects and assumed each laser oscillating in a single longitudinal mode with the same resonance frequency and the same pump current level. Effect of disorder in resonance frequencies will be briefly considered in Sec.IV. Mode coupling is rather generally non-Hermitian, i.e. it corresponds to $\kappa_{n,l} \neq \kappa_{l,n}^*$ for some index $n \neq l$. For dissipative coupling, i.e. if mode coupling is realized without amplifying elements, any eigenvalue of the matrix $\{ \kappa_{n,l} \}$ has negative (for dissipative coupling) or vanishing (for conservative coupling) real part. Dissipative coupling arises rather generally when using diffractive coupling methods \cite{r51,r54,r57,r59,r60,r63,r66bis,r67,r68}, i.e. non-local coupling methods, such as those based on Talbot cavities \cite{r59,r63,r68bis} and diffractive optics. However, dissipative coupling can arise also via evanescent mode coupling, i.e. for local (nearest-neighbor) array coupling, in the presence of dissipative dielectrics, as discussed in Refs.\cite{r33,r74,r75,r76,r77,r78,r79,r80,r81}.  We note that a rather flexible method to tailor coupling constants $\kappa_{n,l}$ in a reconfigurable way has been suggested and experimentally demonstrated in a recent work \cite{r68}. In the following analysis, we will assume discrete rotational invariance along the ring, so that the coupling matrix element $\kappa_{n,l}$ depends on the index difference $(l-n)$ solely, i.e.
\begin{equation}
\kappa_{n,l}= \kappa_{l-n}.
\end{equation}
For $l=n$, without loss of generality the self-coupling term $\kappa_{0}$ can be assumed to be imaginary, i.e. $\kappa_0=-i \gamma$ with $\gamma  \geq 0$ for dissipative coupling: the term $\gamma$ basically describes extra linear loss in each laser of the array arising from the coupling. Under such assumptions, the rate equations (1) and (2) take the form
\begin{eqnarray}
\frac{dE_n}{dt} & = & (1-i \alpha)Z_nE_n-\gamma E_n -i \sum_{\sigma \neq 0} \kappa_{\sigma}E_{n+ \sigma} \\
T \frac{dZ_n}{d t} & = & p-Z_n-(1+2Z_n)|E_n|^2
\end{eqnarray}   
with $\kappa_{-\sigma}=\kappa_{\sigma}^*$ and $\gamma=0$ in the limiting case of Hermitian coupling.
Equations (4) and (5) should be supplemented with the ring (periodic) boundary conditions
\begin{equation}
E_{n+N}(t)=E_n(t).
\end{equation}
\subsection{Stationary phase-locked laser supermodes}
The laser equations (4) and (5) can display different types of stationary states and their dynamics depends largely on the coupling
topology and the strength of coupling between the individual
lasers.  The simplest family of stationary states, corresponding to lasing states in the various supermodes of the ring, is given by \cite{r64,r71}
\begin{equation}
E_{n}^{(st)}(t)=A \exp(iqn+i \omega t) \; , \;\; Z_n^{(st)}(t)=Z
\end{equation}
where
\begin{eqnarray}
Z & = & \gamma- {\rm Im} \left( \sum_{\sigma \neq 0} \kappa_{\sigma} \exp(i q \sigma) \right) \\
\omega & = & - \alpha Z- {\rm Re} \left( \sum_{\sigma \neq 0} \kappa_{\sigma} \exp(iq \sigma)  \right) \\
A & = & \sqrt{\frac{p-Z}{1+2Z}}.
\end{eqnarray}
In the above equations, $q$ is the Bloch wave number of the supermode, which is quantized and can assume $N$ values according to the ring boundary conditions (6)
\begin{equation}
q=q_l= \frac{2 \pi l}{N}
\end{equation}
($l=0,2,...,N-1$). For a sufficiently large value of $N$, the Bloch wave number $q$ can be basically considered a continuous variable, whereas for a small number of lasers in the ring finite size effects should be properly considered; in particular some differences occur for odd and even values of $N$ \cite{r64}. In our work, we will typically assume a sufficiently large number $N$ of lasers, so that $q$ can be treated as an almost continuous variable, and do not consider distinctions between odd and even numbers of lasers.  
Note that the existence domain of the supermode with Bloch wave number $q$ is defined by the inequality $p \geq Z$, so that the excess pump current thresholds of various Bloch supermodes are given by
\begin{equation}
p^{(th)}(q)=\gamma-{\rm Im} \left( \sum_{\sigma \neq 0} \kappa_{\sigma} \exp(iq \sigma)  \right).
\end{equation}
For a dissipative coupling one has $p^{(th)}(q) \geq 0$. Note that in the limiting case of Hermitian coupling, i.e. for $\gamma=0$ and $\kappa_{-\sigma}=\kappa_{\sigma}^*$, one has $p^{(th)}(q)=0$ independent of $q$, i.e. all supermodes are degenerate in threshold. On the other hand, for non-Hermitian coupling the threshold value of injection current depends on $q$, and a supermode with the lowest current threshold is rather generally found. We will specifically focus our analysis to three coupling configurations, corresponding to nearest-neighbor mode coupling.\par
{\it 1. Hermitian coupling.} This coupling corresponds to $\kappa_{\sigma}=0$ for $\sigma \neq \pm 1$ and $\kappa_{-1}= \kappa_1 \equiv \kappa$ real and positive. This case describes the ordinary Hermitian (conservative) mode coupling of nearest-neighbor lasers in the array, which was previously studied in Ref.\cite{r64}. In this case one has:
\begin{equation}
Z=0 \; , \; \; \omega=-2 \kappa \cos (q) \; ,\;\; A= \sqrt{p}.
\end{equation}  
All supermodes have the same threshold value $p^{(th)}=0$.\par
{\it 2. Type-I non-Hermitian coupling.} This case corresponds to $\kappa_{\sigma}=0$ for $\sigma \neq \pm 1$, $\kappa_{-1}=\kappa_1= \kappa_R+i \kappa_I$ and $\gamma=2 \kappa_I$, where $\kappa_R>0$ and $\kappa_I>0$ describe conservative and dissipative couplings between nearest neighbor lasers in the array.  The Hermitian coupling is obtained in the limit $\kappa_I=0$. This kind of non-Hermitian coupling in quite common in coupled laser arrays, and has been considered in some previous works \cite{r75,r76,r81}, especially for two coupled semiconductor lasers \cite{r75,r76}. In this case one has
\begin{eqnarray}
Z & = & 2 \kappa_I [1-\cos(q)] \\ 
\omega & = & - 2 \alpha \kappa_I [1-\cos(q)] -2 \kappa_R \cos (q) \\
A & = & \sqrt{\frac{p-2 \kappa_I [1- \cos(q)]}{1+4 \kappa_I [1- \cos (q)] }}.
\end{eqnarray}
The threshold value of the various supermodes is given by
\begin{equation}
p^{(th)}(q)=2 \kappa_I [1- \cos(q)].
\end{equation}
Note that the supermode with the lowest threshold $p^{(th)}=0$ is the one with $q=0$, i.e. with all lasers in the ring oscillating with the same phase. \par
{\it 3. Type-II non-Hermitian coupling.} This case corresponds to $\kappa_{\sigma}=0$ for $\sigma \neq \pm 1$, $\kappa_{-1}=\kappa \exp(-h)$ and $\kappa_1= \kappa \exp(h)$, with $\kappa$ and $h$ real and positive. Note that the limiting case of Hermitian coupling is obtained for $h=0$. This kind of non-Hermitian mode coupling has been recently introduced in Refs.\cite{r33,r46,r82,r83} and non-Hermiticity arises here from the application of an imaginary gauge field (a complex Peierls$^{\prime}$ phase $h$) in the coupling constant $\kappa$. A possible physical implementation of an imaginary gauge field in coupled microring lasers, based on the use of anti resonant link rings with dissipation,  is discussed in \cite{r33,r46,r83}. For this coupling scheme one has
\begin{eqnarray}
Z & = &  2 \kappa \sinh(h)  [1-\sin(q)] \\ 
\omega & = & - 2 \alpha \kappa \sinh (h) [1-\sin(q)] -2 \kappa \cosh (h) \cos (q) \;\; \\
A & = & \sqrt{\frac{p-2 \kappa \sinh(h)  [1- \sin(q)]}{1+4 \kappa \sinh (h) [1- \sin (q)] }}.
\end{eqnarray}
The threshold value of the various supermodes is given by
\begin{equation}
p^{(th)}(q)=2 \kappa \sinh (h)  [1- \sin(q)].
\end{equation}
Note that the supermode with the lowest threshold $p^{(th)}=0$ is the one with $q=\pi/2$. For such a supermode, the far-field emitted beam carries a non-vanishing orbital angular momentum, i.e. a topological charge \cite{r73}. \par
The behavior of the excess pump current threshold curves $p^{(th)}$ and frequency $\omega$ of the laser array supermodes, versus the Bloch wave number $q$, for the three kinds of nearest-neighbor coupling schemes discussed above is shown in Fig.1(c).\par
We note that type-I and type-II non-Hermitian couplings can be regarded as special cases of nearest-neighbor couplings with arbitrary (complex) values of $\kappa_{-1}$, $\kappa_{1}$, with $\kappa_{-1} \neq \kappa_{1}^{*}$. While the present analysis could be readily extended to include such a more general case, here we limit ourselves to consider type-I and type-II couplings, which are the more common type of nearest-neighbor couplings in lasers. Finally, it should be noted that in case of global coupling other kinds of solutions to the laser equations (4-6) can be found, such as splay states and chimera states (i.e. coexisting synchronous and desynchronous oscillatory behavior) \cite{r71,r84,r85}. Recently, chimera states in nearest-neighbor coupling semiconductor lasers with Hermitian coupling and frequency detuning have been studied in Ref.\cite{r70bis}. However, in this work we will not consider such a type of solutions and their stability. While they are interesting from the viewpoint of complex dynamical systems and networks, in practical cases one should avoided them and phase-locked states, with all lasers emitting in a  synchronous way, are desired.

\begin{figure*}[htbp]
 \includegraphics[width=140mm]{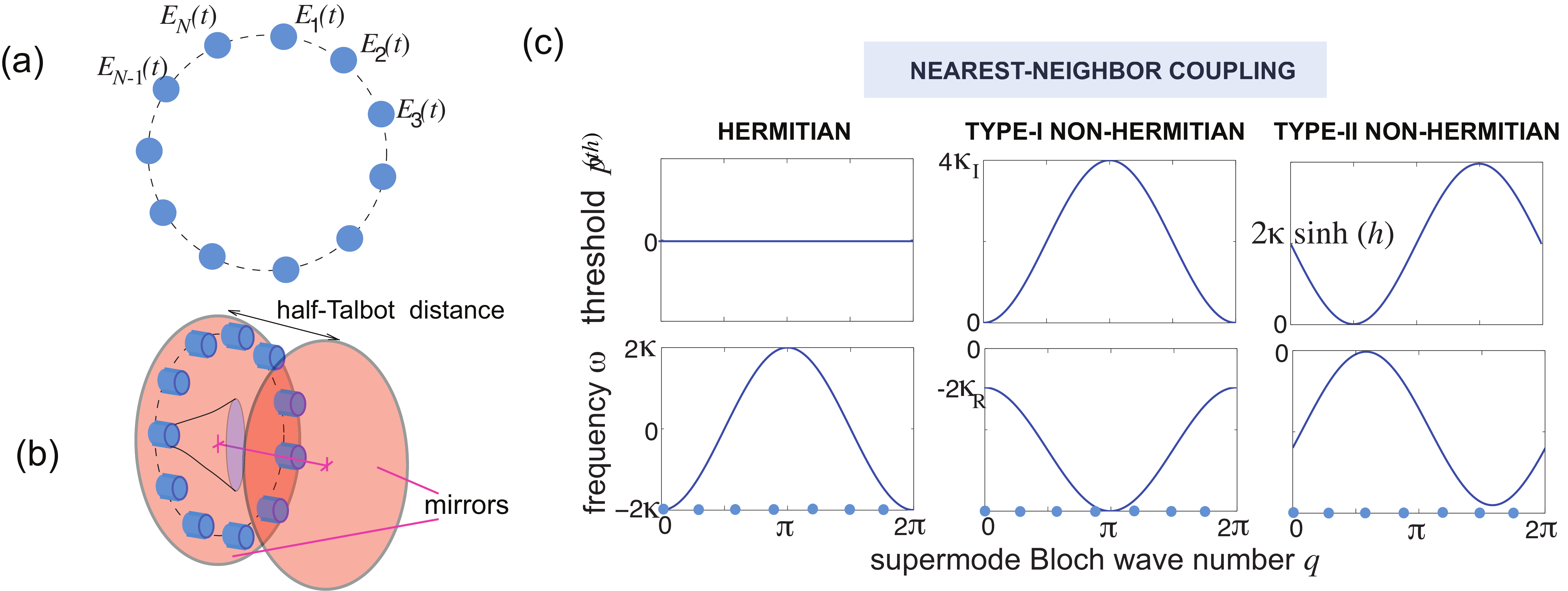}\\
  \caption{(a) Schematic of an array made of $N$ coupled semiconductor lasers on a ring. The coupling can be either local, throughout evanescent mode coupling (nearest-neighbor coupling), or global via some external cavity [for example diffractive coupling in a Talbot cavity, as shown in panel (b)]. Discrete rotational invariance along the ring is assumed. This means that the matrix of coupling constants  $\kappa_{n,l}$ is a function of index difference $(l-n)$ solely, i.e. $\kappa_{n,l}=\kappa_{l-n}$. Dissipative coupling makes the coupling matrix non-Hermitian. (c) Typical behavior of normalized excess pump current threshold $p^{(th)}$ and oscillation frequency $\omega$ of stationary array supermodes versus Bloch wave number $q$ for three kinds of nearest-neighbor couplings: Hermitian coupling $\kappa_{-1}=\kappa_1= \kappa$, with $\kappa$ real positive (left panels); type-I non-Hermitian coupling $\kappa_{-1}=\kappa_1=\kappa_R+i \kappa_I$, with $\kappa_R$, $\kappa_I$ real and positive (central panels); type-II non-Hermitian coupling $\kappa_{-1}=\kappa \exp(-h)$, $\kappa_1= \kappa \exp(h)$ with $\kappa$, $h$ real and positive (right panels). The Bloch wave number $q$ of supermodes is quantized according to Eq.(11) given in the text and can assume $N$ values. }
\end{figure*}

\section{Linear stability analysis}
Dynamical instabilities in arrays of coupled semiconductor lasers are known to arise in a wide range of parameter operations corresponding to realistic conditions 
\cite{r58,r61,r62,r64,r71}, even when delayed coupling and frequency detuning effects are negligible. In particular, a detailed analysis of the instability arising in a ring geometry with nearest-neighbor Hermitian coupling has been presented by Li and Erneux in Ref.\cite{r64} (see also \cite{r71}). A natural question then arises: what is the impact of non-Hermitian coupling on the onset of dynamical instabilities? Can non-Hermitian coupling help to prevent laser instabilities and force stable laser emission in the preferred supermode with $q=0$ (all lasers in the array emitting in phase) or in a supermode that corresponds to a vortex beam in far-field? It is clear that some non-local coupling methods known in the literature, such as those based on the Talbot effect, can be regarded as a kind of non-Hermitian coupling scheme \cite{note1}  and they help to achieve stable laser emission. To study the impact of non-Hermitian coupling on laser instabilities in a rather general framework, we performed a detailed linear stability analysis of the phase-locked solutions given by Eqs.(7-10), extending the analysis of Ref.\cite{r64} to account for a rather broad class of non-Hermitian coupling configurations. As we will see, even for nearest-neighbor coupling non-Hermitian effects can effectively suppress the onset of dynamical instabilities and enable stable phase locking operation in a supermode with either $q=0$ (for type-I non-Hermitian coupling) or $q=\pi/2$ (for type-II non-Hermitian coupling). After setting $E_n(t)=E_{n}^{(st)} (t) [1+ \delta E_n (t)]$ and $Z_n(t)=Z [1+ \delta Z_n(t)]$, the linearized equations that describe the evolution of small perturbations $\delta E_n(t)$ and $\delta Z_n$ from the stationary state read
\begin{eqnarray}
\frac{d \delta E_n}{dt} & = & (1-i \alpha) \delta Z_n \nonumber \\
& - & i \sum_{\sigma \neq 0} \kappa_{\sigma} \exp(iq \sigma) \left( \delta E_{n+ \sigma} -\delta E_n \right)  \\
T \frac{d \delta Z_n}{dt} & = & -(1+2A^2) \delta Z_n \nonumber \\
&  - & A^2 (1+2Z) \left( \delta E_n+\delta E_n^* \right)
\end{eqnarray}
with the periodic ring boundary conditions $\delta E_{n+N}(t)=\delta E_n(t)$. The most general solution to Eqs.(22) and (23) is a linear superposition of solutions of the form
\begin{eqnarray}
\delta E_n(t) & = & R_1 \exp(i Qn+\lambda t)+R_2^* \exp(-iQn+ \lambda^* t) \;\;\;\;\; \\
\delta Z_n(t) & = & P \exp(i Qn+\lambda t)+P^* \exp(-iQn+ \lambda^* t) \;\;
\end{eqnarray}
where $Q$ is the Bloch wave number of the perturbation [quantized like $q$ according to Eq.(11)] and $\lambda$ describes the growth rate of the perturbation. The complex amplitudes $R_1$, $R_2$ and $P$ satisfy the homogeneous linear system
\begin{eqnarray}
\lambda R_1 & = & (1-i \alpha)P-i \theta_1 R_1 \\
\lambda R_2 & = & (1+i \alpha) P+ i \theta_2 R_2\\
T \lambda P & = & -(1+2A^2)P-A^2(1+2Z)(R_1+R_2)
\end{eqnarray}
where we have set
\begin{eqnarray}
\theta_1 & \equiv & \sum_{\sigma \neq 0}  \kappa_{\sigma} \exp(i q \sigma) \left[ \exp(i Q \sigma) -1 \right]   \\
\theta_2 & \equiv & \sum_{\sigma \neq 0}  \kappa_{\sigma}^* \exp(-i q \sigma) \left[ \exp(i Q \sigma) -1 \right].
\end{eqnarray}
The growth rate $\lambda$ is obtained from the corresponding eigenvalue problem, i.e. $\lambda$ is a root of the cubic equation
\begin{equation}
\lambda^3+c_1 \lambda^2+ c_2 \lambda+c_3=0
\end{equation} 
where we have set
\begin{eqnarray}
c_1 & \equiv & i (\theta_1-\theta_2)+\frac{1+2A^2}{T} \\
c_2 & \equiv & \theta_1 \theta_2+ \frac{i(\theta_1-\theta_2)(1+2A^2)+2A^2(1+2Z)}{T} \\
c_3 & \equiv & \frac{\theta_1 \theta_2 (1+2A^2)}{T } \nonumber \\
& + & \frac{A^2(1+2Z)[i (\theta_1-\theta_2)-\alpha( \theta_1+\theta_2)]}{T}.
\end{eqnarray}
Note that, for a given value of the Bloch wave number $q$ of stationary array supermode, one has three possible values $\lambda=\lambda_l(Q)$ ($l=1,2,3$) of the perturbation growth rate, which depend on the Bloch wave number $Q$ of the perturbation. The stationary phase-locked supermode with Bloch wave number $q$, given by Eqs.(7-10), is thus linearly stable provided that the real part of any of the three eigenvalue $\lambda_l(Q)$  is positive or vanishing, for {\it any} wave number $Q$ of the perturbation. Owing to phase invariance of the stationary state solution, one of the three eigenvalue vanishes at $Q=0$. The roots of the cubic equation (31) are given in the most general case by Cardano$^{\prime}$ formula, however their form is rather cumbersome to be given here and in general one has to resort to a numerical computation of the eigenvalues and corresponding domain of stability. Some analytical insights can be obtained under proper scaling of parameters, as suggested in Ref.\cite{r64}. Taking into account that in a semiconductor laser $T$ is a large parameter ($T \sim 100-1000$), we may introduce a small parameter $\epsilon$ defined by $\epsilon=1/ \sqrt{T}$ and find the roots of Eq.(31) as a power series in $\epsilon$. Moreover, since the instability arises for a strength of coupling constants of order $\sim \epsilon^2$ \cite{r64}, we assume $\kappa_{\sigma}$ small and of order $\sim \epsilon^2$, i.e. we set $\kappa_{\sigma} \equiv \epsilon^2 \beta_{\sigma}$, with $\beta_{\sigma} \sim O(1)$. With such a scaling, one has $c_1 \sim \epsilon^2$, $c_2 \sim \epsilon^2$ and $c_3 \sim \epsilon^4$. We then look for a solution to the cubic equation (31) in power series of $\epsilon$, namely we assume
\begin{equation}
\lambda= \epsilon (\lambda_0 + \epsilon \lambda_1 +...).
\end{equation}
At leading order in $\epsilon$, the three roots of the cubic equation are found to be given by
\begin{eqnarray}
\lambda_1& = &- \frac{c_3}{c_2}+o(\epsilon^2) \\
\lambda_{2,3} & = & \frac{c_3-c_1c_2}{2 c_2} \pm i \sqrt{c_2} +o (\epsilon^2)
\end{eqnarray}
with $c_1=i (\theta_1-\theta_2)+(1+2A^2)/T$, $c_2 \simeq 2A^2(1+2Z)/T$ and $c_3 \simeq (c_2/2) [i(\theta_1-\theta_2)-\alpha(\theta_1+\theta_2))]$.
The stability condition, ${\rm Re}(\lambda_{1,2,3}) \leq 0$, then yields
\begin{equation}
0 \leq {\rm Re}(c_3) \leq c_2 {\rm Re}(c_1)
 \end{equation}
Substitution of Eqs.(32-34) into Eq.(38) and using Eqs.(29) and (30) finally yields the following stability conditions at leading order in $\epsilon$
\begin{eqnarray}
\sum_{\sigma \neq 0} \left[ 1-\cos(Q \sigma) \right] \left[ (\alpha-i)  \kappa_{\sigma} \exp(i q \sigma) + c.c. \right] \geq 0 \;\;\;\;\;\;\;\; \\
\frac{1}{2} \sum_{\sigma \neq 0}  \left[ 1-\cos(Q \sigma) \right] \left[ (\alpha-i)  \kappa_{\sigma} \exp(i q \sigma) + c.c. \right] \nonumber \\
 \leq \frac{1+2A^2}{T}-\sum_{\sigma \neq 0}  \left[ 1-\cos(Q \sigma) \right] \left[ i  \kappa_{\sigma} \exp(i q \sigma) + c.c. \right] \;\;\;\;\;
\end{eqnarray}
\begin{figure*}[htbp]
 \includegraphics[width=140mm]{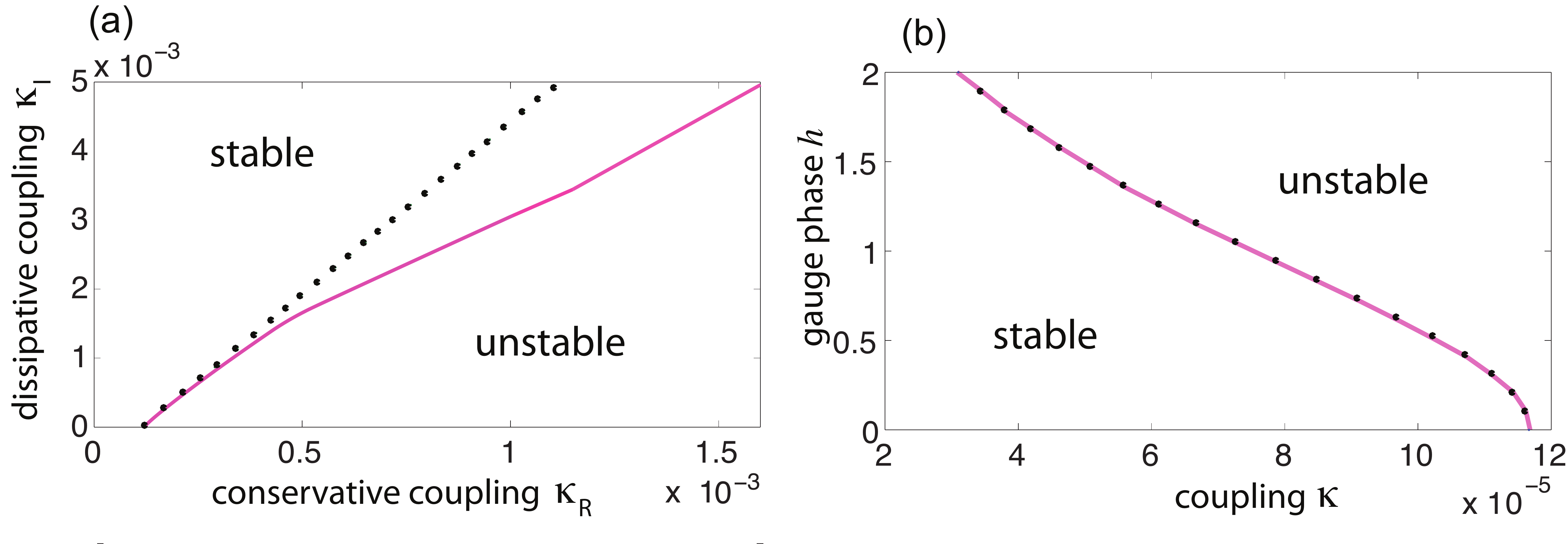}\\
  \caption{(color online) Numerically-computed stability diagram of the in-phase supermode state ($q=0$)  (a) for type-I non-Hermitian coupling ($\kappa_{-1}=\kappa_1=\kappa_R+i \kappa_I$) in the plane $(\kappa_R,\kappa_I)$ of conservative/dissipative coupling strengths, and (b) for type-II non-Hermitian  coupling ($\kappa_{-1}=\kappa \exp(-h)$, $\kappa_1=\kappa \exp(h)$) in the $(\kappa,h)$ plane. Parameter values are $\alpha=5$, $T=600$ and $p=0.2$. Dotted curves refer to the stability boundaries as obtained from the asymptotic analysis of the roots of the cubic determinantal equation (31). The Hermitian limit is retrieved for $\kappa_I=0$ in (a), and $h=0$ in (b). In this case the Hopf instability arises for a coupling strength larger than $\kappa^{(max)} \simeq 1.167 \times 10^{-4}$, given by Eq.(43) with $q=0$.}
\end{figure*}
which should be satisfied for any wave number $Q= 2 \pi l/N$ ($l=0,1,2,...,N-1$) of perturbation. The stability conditions (39) and (40) apply to a rather arbitrary coupling scheme, i.e. either local or global couplings, with the 
solely constraint of translational invariance. Let us now specialize the general results to the three local (nearest-neighbor) coupling schemes introduced in the previous section [Fig.1(c)].\par
{\it 1. Hermitian coupling.} This case was considered in Ref.\cite{r64} and corresponds to $\kappa_{\sigma}=0$ for $\sigma \neq \pm 1$ and $\kappa_1=\kappa_{-1}=\kappa$ real and positive. In this case conditions (39) and (40) reads explicitly
\begin{eqnarray}
\cos (q)  \geq 0 \\
\kappa  \leq  \frac{1+2p}{2 \alpha T \cos (q) [1-\cos (Q)]}
\end{eqnarray}
which have been previously derived in Ref.\cite{r64}. Equation (41) indicates that only the supermodes with Bloch wave number $q$ in the range $|q|< \pi/2$ are stable states, whereas Eq.(42) shows that a Hopf instability at the frequency $\omega_{H}= \sqrt{c_2}= \sqrt{2p/T}$ arises for large enough coupling constant $\kappa$ [violation of Eq.(42) corresponds to the two complex-conjugate eigenvalues $\lambda_{2,3}$, given by Eq.(37),  to become unstable]. Clearly, the most unstable perturbation for the emergence of the Hopf instability is the one with Bloch wave number $Q= \pi$, and the maximum value of coupling constant, below which the phase-locked supermode with wave number $q$ remains stable, is given by
\begin{equation}
\kappa^{(max)}=\frac{1+2p}{4 \alpha T\cos (q)}
\end{equation}
Note that the most unstable supermode is the one with $q=0$, i.e. the supermode with in-phase laser emission, as previously shown in Ref.\cite{r64}. \par
{\it 2. Type-I non-Hermitian coupling.} Let us assume $\kappa_{-1}=\kappa_1 \equiv \kappa_R+i \kappa_I$, where $\kappa_{R}>0$ and $\kappa_I>0$ are the conservative and dissipative couplings, and $\kappa_{\sigma}=0$ for $|\sigma|>1$. In this case the stability conditions (39) and (40) read explicitly
\begin{eqnarray}
(\kappa_R \alpha + \kappa_I) \cos (q)  \geq  0 \\
\kappa_R \alpha-\kappa_I  \leq \frac{1+2A^2}{2 T \cos (q) [1-\cos (Q)]}.
\end{eqnarray}
Like in the Hermitian case discussed above, Eq.(44) shows that the supermodes with $\cos (q)< 0$ are always unstable, while stable supermodes necessarily should correspond to a Bloch wave number $q$ with $\cos (q) \geq 0$. The main impact of non-Hermitian coupling is clear when considering the stability condition (45). Remarkably, for a sufficiently large value of the dissipative coupling term as compared to the conservative one, namely for
\begin{equation}
\kappa_I \geq \alpha \kappa_R
\end{equation} 
Eq.(45) is satisfied for $\cos (q)>0$, regardless of the strength of the couplings $\kappa_I$ and $\kappa_R$. This means that, provided that Eq.(46) is satisfied, non-Hermitian coupling can prevent the onset of the Hopf instability observed in the Hermitian limit as the coupling strength between neighboring lasers is increased. \par
{\it 3. Type-II non-Hermitian coupling.} Let us assume $\kappa_{-1}=\kappa \exp(-h)$ and $\kappa_1= \kappa \exp(h)$, with $\kappa$ and $h$ real and positive constant, and $\kappa_{\sigma}=0$ for $|\sigma|>1$. In this case the stability conditions (39) and (40) read explicitly  
\begin{eqnarray}
\alpha \cosh (h) \cos (q)+\sinh(h) \sin (q) >0 \;\;\;\;\; \\
 \kappa \left[ \alpha \cosh(h) \cos (q)- \sinh(h) \sin (q) \right]  \nonumber \\
 < \frac{1+2A^2}{2T[1-\cos(Q)]} \;\;\;\;\;
\end{eqnarray}

\begin{figure*}[htbp]
 \includegraphics[width=140mm]{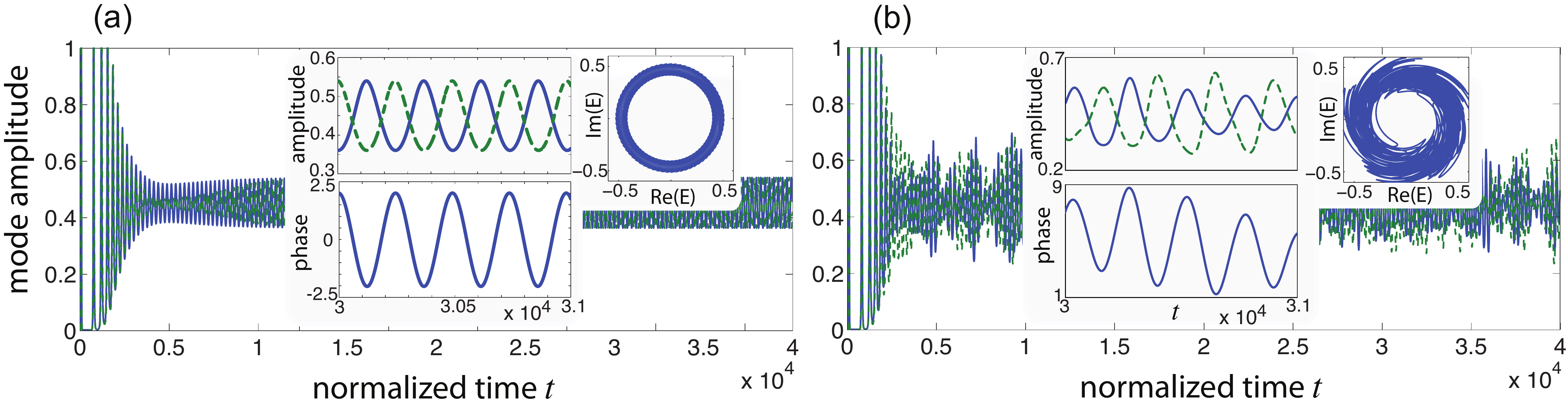}\\
  \caption{Numerically-computed laser switch on dynamics in a ring array made of $N=8$ lasers for nearest-neighbor Hermitian coupling in the oscillatory (Hopf) instability regime. Parameter values are $T=600$, $p=0.2$, $\alpha=5$, and $\kappa=2 \times 10^{-4}$ in (a), $\kappa= 1.5 \times 10^{-3}$ in (b). Initial condition is a small random noise of the field amplitudes $E_n$, and stationary values $Z_n=p$ of normalized excess carriers. The figure shows the behavior of modal amplitude $|E_n|$ for the two modes $n=1$ (solid curve) and $n=4$ (dashed curve) of the array. Insets: upper left inset shows the detailed behavior of the mode amplitudes after relaxation oscillation transient; lower left inset shows the behavior of the phase difference between the two modes; right inset depicts the phase space evolution of field amplitude $({\rm Re}(E), {\rm Im}(E))$ of the $n=1$ laser in the array, after initial relaxation oscillation transient.}
\end{figure*}

\begin{figure*}[htbp]
 \includegraphics[width=180mm]{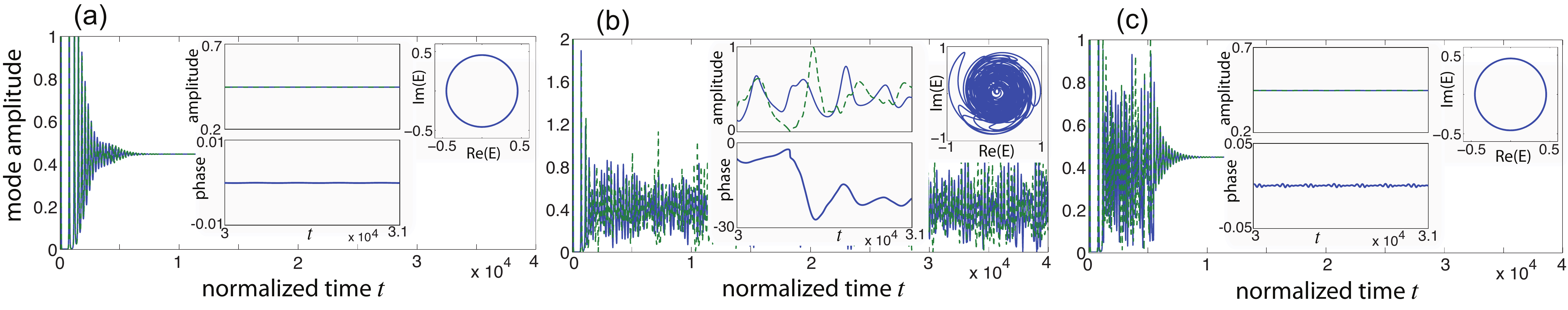}\\
  \caption{Laser switch on dynamics for type-I non-Hermitian coupling. (a) $\kappa_R= 2 \times 10^{-4}$, $\kappa_I= \alpha \kappa_R$. (b) $\kappa_R= 1.5 \times 10^{-3}$ and $\kappa_I= \alpha \kappa_R$. (c) $\kappa_R= 1.5 \times 10^{-3}$ and $\kappa_I= 4 \alpha \kappa_R$. Other parameter values are as in Fig.3 ($T=600$, $p=0.2$, $\alpha=5$).}
\end{figure*}

\begin{figure*}[htbp]
  \includegraphics[width=180mm]{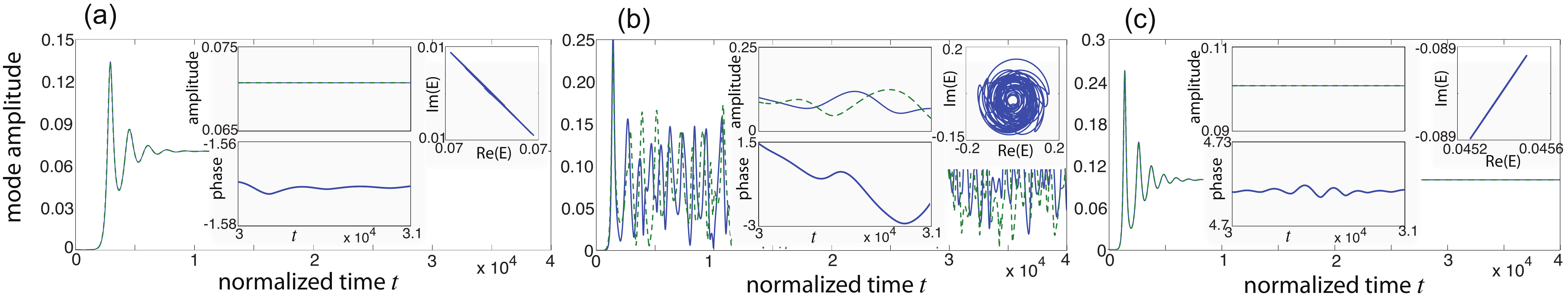}\\
  \caption{Laser switch on dynamics for type-I non-Hermitian coupling. Parameter values are: (a) $\kappa=2 \times 10^{-4}$, $h=3$, $p=0.005$; (b)
   $\kappa=2 \times 10^{-4}$, $h=3$, $p=0.01$; (c)  $\kappa=2 \times 10^{-4}$, $h=4$, $p=0.01$. Other parameter values are as in Fig.3 ($\alpha=5$, $T=600$).
  Stable phase-locking laser emission in the $q= \pi/2$ supermode is observed in (a) and (c).}
\end{figure*}

Note that, in this case, the supermode with the lowest current threshold, corresponding to $q=\pi/2$, is always stable, regardless of the strength $\kappa$ of the laser coupling, even for a small value of the non-Hermitian gauge field $h$. Therefore type-II non-Hermitian coupling is expected to be a suitable and robust means to generate stable phase-locked laser emission in a supermode carrying a non-vanishing topological charge. This is because the imaginary gauge field, corresponding to asymmetric coupling $\kappa_{-1} \neq \kappa_1$, introduces a chiral behavior in the dynamics \cite{r28,r29}. On the other, from Eq.(48) it follows that the non-Hermitian gauge field $h$ {\it reduces} the stability domain of the in-phase supermode $q=0$, i.e. it enhances the onset of instability for this supermode at lower values of coupling $\kappa$. This is because the imaginary gauge field introduces a preferred {\it directional} transport along the chain of coupled lasers \cite{r33}. \par
The approximate stability conditions, given by Eqs.(39) and (40) and obtained by an asymptotic form of the roots of the cubic determinantal equation, correctly capture the main role played by non-Hermitian coupling 
in preventing or enhancing the onset of the Hopf instability. As an example, Fig.2 shows the exact numerically-computed stability domains of the in-phase supermode state ($q=0$) for type-I non-Hermitian coupling in the $(\kappa_R,\kappa_I)$ plane, and for type-II non-Hermitian coupling in the $(\kappa,h)$ plane. The exact stability domains are also compared to those obtained by the asymptotic analysis of the eigenvalues of the cubic determinantal equation. Note that, as expected, the asymptotic analysis provides a good approximation of the stability boundaries only for relatively small values of coupling strength. Note also that, as excepted from the asymptotic analysis, while type-I non-Hermitian coupling prevents the onset of instability for the in-phase supermode $q=0$ [Fig.2(a)], type-II non-Hermitian coupling narrows the stability region of this supermode [Fig.2(b)].

\section{Numerical results}
The ability of non-Hermitian couplings to prevent the onset of dynamical instabilities and to force stable laser oscillation in the in-phase ($q=0$) supermode or in a chiral ($q=\pi/2$) supermode has been checked by direct numerical simulations of laser rate equations (1-2). Parameter values used in the simulations are typical of semiconductor laser arrays and comparable to those used in previous theoretical works  \cite{r58,r62,r64,r65}: $\alpha=5$, $T=600$, and $p$ ranging from 0.003 to 0.2. The rate equations have been numerically solved using an accurate variable-step Runge-Kutta method, assuming $N=8$ lasers in the ring. As an initial condition, we typically assumed small random values of amplitudes $E_n$ for the electric fields and the stationary values $Z_n=p$ of excess carrier densities in each laser of the array. After initial relaxation oscillation transient describing laser switch on, different dynamical regimes can be observed, which depend on parameter values but can also depend on initial conditions, i.e. different runs starting from small random noise can result in different dynamical behaviors. This is a clear signature of multi stability and of highly nonlinear dynamics of laser array systems, which is very common in coupled nonlinear oscillator models (see for instance \cite{r86} and references therein).  Here, we are not aimed to provide a comprehensive study of the rich and complex dynamical behavior of the laser array that could be observed in parameter space, rather we want to show how non-Hermitian coupling can suppress dynamical instabilities found for Hermitian coupling, thus proving a possible route for stable high-power laser array design. As an example, Fig.3(a) shows a typical behavior of laser emission started from initial random noise as obtained for nearest-neighbor Hermitian coupling in the Hopf instability region  for a pump parameter $p=0.2$ and for a coupling  constant $\kappa=2 \times 10^{-4}$, which is $\sim 1.71 $ times larger than the maximum value $\kappa^{(max)} \simeq 1.167 \times 10^{-4}$ predicted by the linear stability analysis  [Eq.(43) and Fig.2]. The laser amplitudes undergo self-pulsation as a result of the Hopf instability, at a frequency $\omega_{H} \simeq 243$ which is very close to the theoretical value $\omega_{H}= \sqrt{2 p/T}$ predicted by the linear stability analysis. For Hermitian coupling, more irregular behaviors are observed as the coupling strength $\kappa$ is further increased, as shown for example in Fig.3(b). The suppression of supermode instability and stable oscillation of the array in the $q=0$ supermode for type-I non-Hermitian coupling is shown in Fig.4(a). Parameter values are as in Fig.3(a), except that the couplings $\kappa_1=\kappa_2=\kappa_R+i \kappa_I$ have a non-vanishing dissipative part $\kappa_I$. Note that, for $\kappa_I=\alpha \kappa_R$ chosen in the numerical simulations, according to the linear stability analysis [Eq.(46)] all supermodes of the array are locally stable. Since $q=0$ is the supermode with the lowest pump current threshold, it is thus expected that this is the most rapidly growing mode from initial noise and the stable attractor of the dynamics after laser switching on, as Fig.4(a) indicates. It should be noted that, for a laser well above threshold (e.g. $p=0.2$), increasing further the coupling $\kappa_R$, yet keeping the ratio $\kappa_I / \kappa_R= \alpha$ constant, results in a typical irregular behavior of laser output starting from initial small random noise, as shown in Fig.4(b). Such an irregular behavior does not arise from linear instability of laser supermodes, which are all linearly stable \cite{note2}, rather it is most likely due to highly nonlinear mixing of supermodes, which are all well above threshold, and the filtering effect introduced by the non-Hermitian coupling is not effective in preventing oscillation of supermodes with higher threshold than the $q=0$ supermode \cite{note3}. In this case, to achieve stable phase-locking emission in the $q=0$ supermode one can either decrease the pump current level $p$ \cite{note3} or increase the ratio $\kappa_I / \kappa_R$ of dissipative to conservative coupling terms, which makes the filtering effect stronger [see for instance Fig.4(c)].\\
For type-II non-Hermitian coupling, stable laser oscillation is expected to occur on  the lowest-threshold $q=\pi/2$ supermode, resulting in a far-field vortex beam carrying orbital angular momentum. Figure 5(a) shows an example of stable emission in the $q= \pi/2$ supermode at a relatively low pump current level. Like for type-I non-Hermitian coupling,  the vortex-beam emission is not always the stable attractor of the dynamics when the coupling strength $\kappa$ and/or the pump current level $p$ are increased. Indeed, irregular emission can be observed as well [see Fig.5(b)]. Nevertheless, by increasing the non-Hermitian parameter $h$, so as to filtering out oscillation of other supermodes, or reducing the pump current level closer to the threshold value, one can restore stable emission of the vortex supermode as the laser is switched on, as shown as an example in Fig.5(c).\\
In the previous examples, we assumed that all the lasers oscillate on a single longitudinal mode with the same resonance frequency. However, in practice the lasers can show slight deviations of their resonance frequencies from the ideal one due e.g. to imperfections in fabrication. While deviations of the resonance frequencies much smaller than mode coupling can be neglected, they can destroy phase locking when become comparable to the strength of mode coupling. In case of Hermitian coupling, for a large number $N$ of lasers disorder in the resonance frequencies makes the array supermodes localized rather than extended  (because of Anderson localization), so that independent oscillations in clusters of lasers is observed (see e.g. the recent experiment \cite{r87}). For gradient frequency detunings in the array, complex patterns such as chimera states have been predicted to arise for Hermtian coupling in Ref.\cite{r70bis}.\\
Anderson localization arising from disorder in the resonance frequencies of the lasers occurs as well as for complex but symmetry coupling, i.e. for type-I non-Hermitian coupling. Interestingly, type-II non-Hermitian coupling, corresponding to asymmetric mode-coupling, provides robust chiral transport along the ring, which is immune to moderate disorder strength owing to the phenomenon of non-Hermitian delocalization transition \cite{r88} (see also \cite{r33,r82,r89}): the supermode with the lost threshold $q= \pi/2$ is not localized by moderate disorder of resonance frequencies in the ring. Therefore, we expect that type-II non-Hermitian coupling, besides of combating dynamical instabilities, can ensure stable laser emission even in presence of moderate strength of disorder in the laser resonance frequencies. As an example, Fig.6 compares laser dynamics for type-I and type-II non-Hermitian coupling with the same disorder of the resonance frequencies of the lasers in the ring. Parameter values are the same in the two cases, with comparable strength of mode coupling such that, without disorder, both coupling methods ensures stable phase-locked oscillation [Fig.6(a)]. In the presence of moderate disorder, in type-I non-Hermitian coupling stable oscillation is typically destroyed, with the appearance of irregular oscillations [Fig.6(b), middle panel]. On the other hand, for type-II non-Hermitian coupling phase locking and stable phase-locked laser emission persists despite of disorder [Fig.6(b), right panel].
\begin{figure*}[htbp]
  \includegraphics[width=180mm]{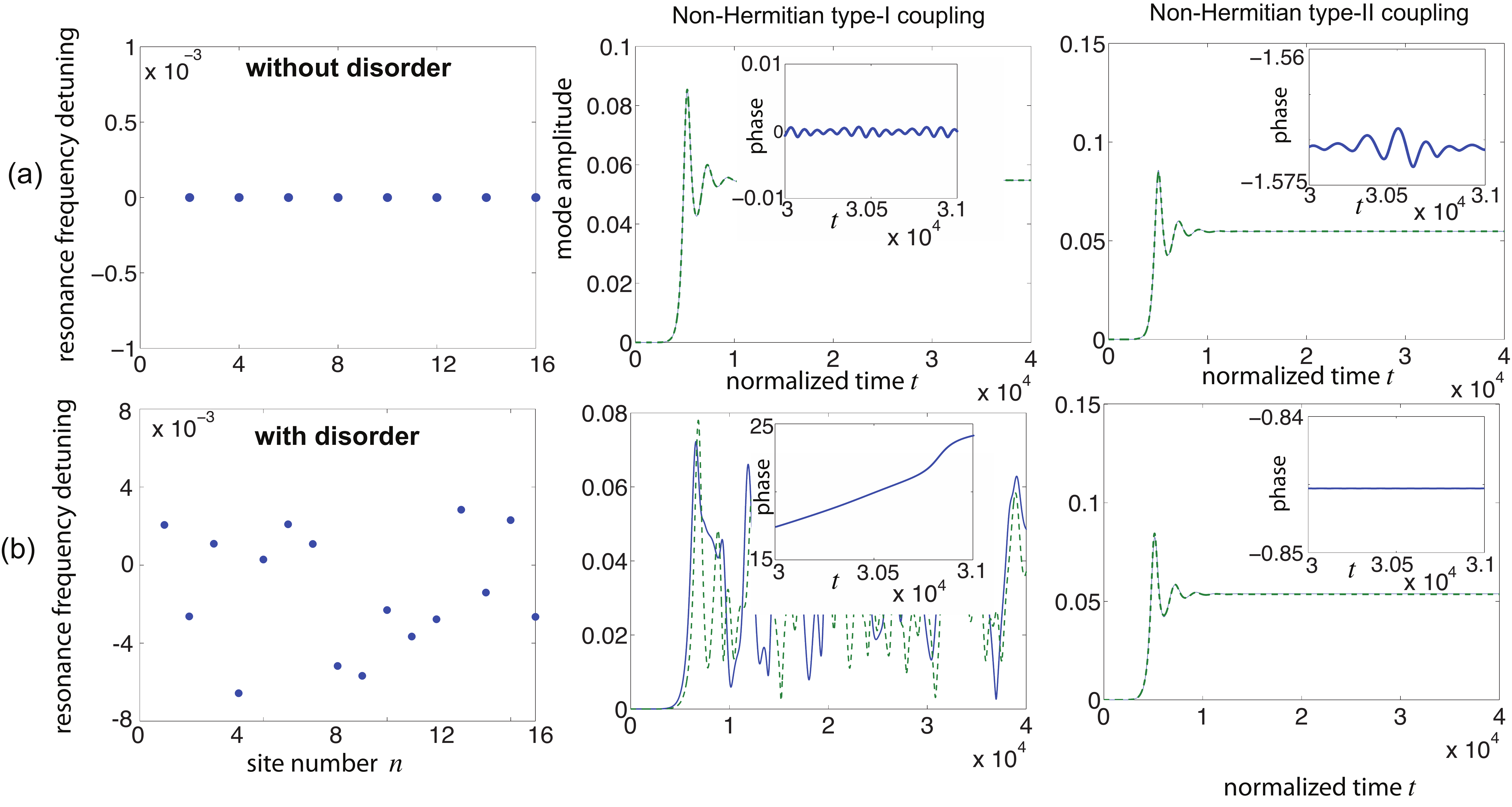}\\
  \caption{Laser switch on dynamics for type-I (middle column) and type-II (right column) non-Hermitian couplings in presence of disorder of laser resonance frequencies for a ring array of $N=16$ coupled lasers. The distribution of the resonance frequency detuning (in units of $ 1 / \tau_p$) is shown in the left column. In (a) there is not disorder, whereas in (b) disorder of resonance frequencies is considered of strength comparable to the coupling constant. In the central and right panels of (a) and (b), solid and dashed curves refer to the field in lasers at sites $n=1$ and $n=8$ of the ring. The insets show the behavior of the phase difference of the fields in the two rings.
  Parameter values are: $\alpha=5$, $T=600$, $p=0.003$ and $\kappa_1=\kappa_2=0.002+0.01i$ for type-I coupling (central column),  $\kappa_1=3.66 \times 10^{-6}$, $\kappa_2=0.0109$ for type-II coupling (right column).}
\end{figure*}

\section{Conclusions}
In this work we have considered the long-standing problem of forcing stable supermode emission in laser arrays in the perspective of the emerging field of non-Hermitian photonics \cite{r1,r2,r4}. Even when time delay effects are negligible, semiconductor laser arrays are known to undergo a great variety of dynamical behaviors, ranging from self-pulsing to chaos, and to show complex spatiotemporal patterns  such as chimera states \cite{r58,r61,r62,r64,r70bis,r71,r84}. While complexity of laser array behavior can be of interest from the viewpoint of the physics of complex systems, combating the onset of dynamical instabilities and forcing synchronous laser emission is desirable in most photonic applications. Using a standard rate equation model describing the dynamics of semiconductor laser arrays on a ring \cite{r64}, we have shown rather generally that non-Hermitian coupling engineering of laser arrays is able to mitigate the onset of dynamical instability and can force laser emission in  a stable supermode. Traditional methods of laser phase locking based on global  couplings, such as diffractive coupling in Talbot cavities, can be regarded as a kind of non-Hermitian coupling engineering.  Here we have shown that non-Hermitian coupling can effectively stabilize laser emission in a given supermode using local (nearest-neighbor) non-Hermitian coupling schemes. In particular, we considered two kinds of non-Hermitian local couplings, referred to as type-I and type-II non-Hermitian couplings.  In the former case all the lasers oscillate with the same phase, whereas in the latter case  $\pi/2$ phase slips between adjacent lasers can be realized, resulting in a far-field vortex beam emission carrying orbital angular momentum. As compared to type-I non-Hermitian coupling, type-II non-Hermitian coupling realizes a chiral transport along the ring which is robust against moderate disorder of resonance frequencies. Our results show that the emerging field of non-Hermitian photonics can find important application into a rather old problem of laser science and technology and are expected to stimulate further theoretical and experimental studies. In particular, type-II non-Hermitian coupling provides a promising scheme in laser array design for combating the detrimental effects of laser instabilities as well as unavoidable disorder of resonance frequencies due to fabrication imperfections.

\acknowledgments 
S.L. acknowledges useful discussions with D. Gomila and I. Fischer. Hospitality at the IFISC
(CSIC-UIB), Palma de Mallorca is also gratefully acknowledged.


\begin{thebibliography}{9}
\bibitem{note1}
For laser arrays with diffraction coupling in a Talbot cavity, the form of the non-Hermitian coupling matrix $\{ \kappa_{n,l} \}$ is given in Ref.\cite{r63}. See also Ref.\cite{r66bis}.
\bibitem{note2}
We numerically checked that, for $\kappa_I= \alpha \kappa_R$ and as $\kappa_R$ is increased, all supermodes of the array remain locally stable when the initial condition is assumed to be a slightly-perturbed array supermode.
\bibitem{note3}
This is suggested by the fact that, decreasing the pump current level from $p=0.2$ to below $p \simeq 0.04$, starting from small random noise amplitudes the final attractor of the dynamics is the $q=0$ supermode, like in Fig.5.
\end{thebibliography}
\end{document}